\documentclass{jpsj-suppl}
\usepackage{graphicx}
\usepackage{epsf}

\title{Compatibility of recent ${\cal S}=-2$ emulsion events} 
\author{Avraham \textsc{Gal}} 
\inst{Racah Institute of Physics, The Hebrew University, Jerusalem 91904, 
ISRAEL} 

\recdate{January 19, 2026} 

\abst{We question the compatibility of recent ${\cal S}=-2$ hypernuclear 
assignments of J-PARC E07 $\Xi^-$-capture emulsion events with assignments 
deduced from other experiments.} 

\kword{strangeness $-2$ nuclear physics} 

\begin{document}

\maketitle

\section{Implications of the J-PARC E05 ${^{12}}{\rm C}(K^-,K^+)$ spectrum} 

\begin{figure}[!ht] 
\begin{center} 
\includegraphics[scale=0.40]{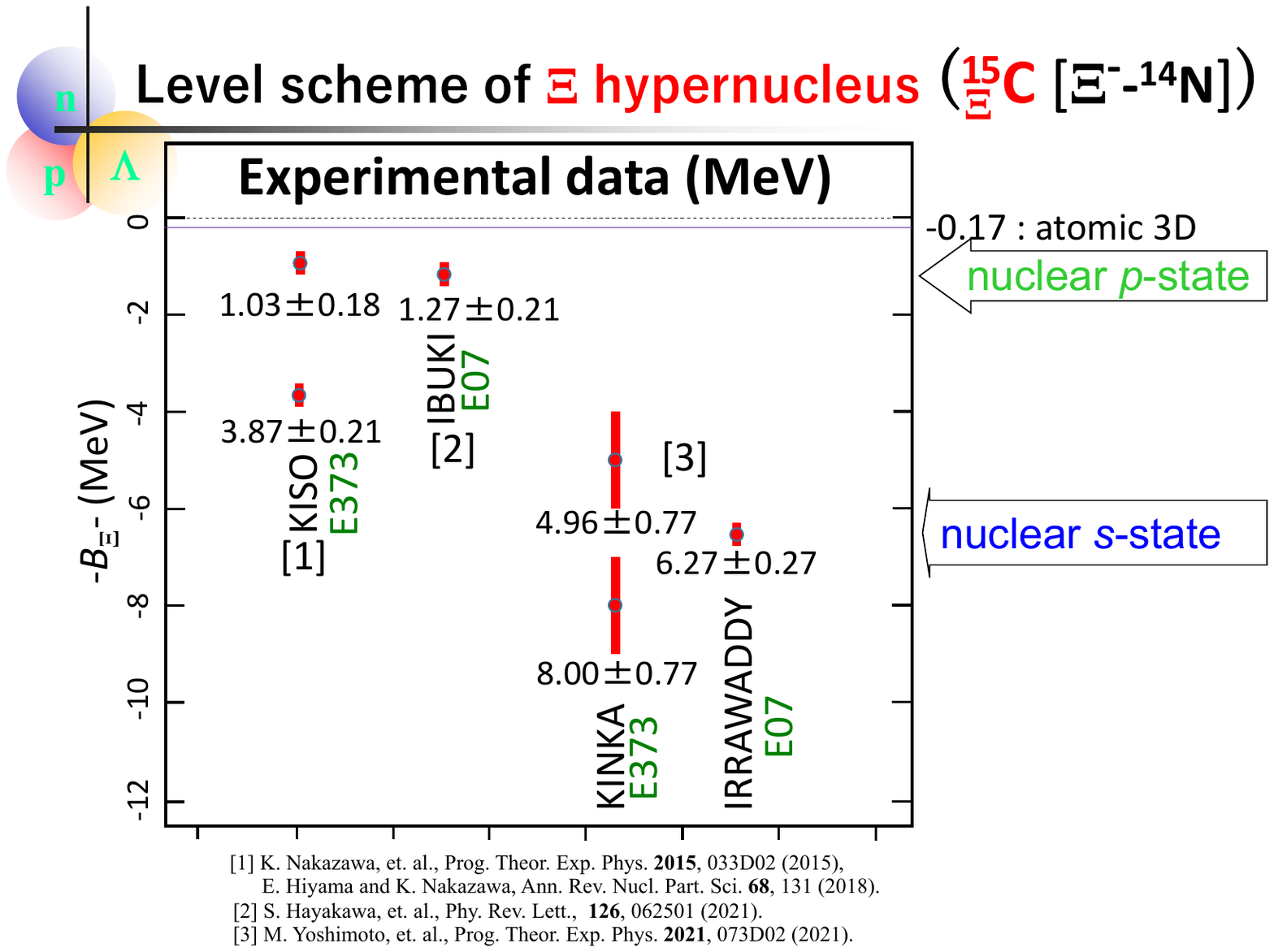} 
\caption{$\Xi^{-}-{^{14}{\rm N}}$ spectrum deduced from $\Xi^-$ capture events 
identified by their twin-$\Lambda$ hypernuclear decays in KEK-PS E373 and 
J-PARC E07 emulsion experiments. Figure adapted from Ref.~\cite{E07}.} 
\label{fig:14N} 
\end{center} 
\end{figure}

J-PARC E05 published recently 
a ${^{12}}{\rm C}(K^-,K^+){^{12}_{\Xi}}{\rm Be}$ spectrum that 
indicates a nuclear $\Xi^-_{1s}-{^{11}{\rm B}}$ state bound by 
$B_{\Xi^-}=8.9\pm 1.4{^{+3.8}_{-3.1}}$~MeV~\cite{E05}. A natural question to 
ask is whether this $\Xi^-_{1s}$ assignment is compatible with nuclear 
$\Xi^-_{1s}-{^{14}{\rm N}}$ assignments of $\Xi^-$-capture emulsion events 
shown in Fig.~\ref{fig:14N}. Indeed, expecting $B_{\Xi^-}$ to increase with 
$A$, it is puzzling to have a $\Xi^-_{1s}-{^{11}{\rm B}}$ binding energy 
larger than a $\Xi^-_{1s}-{^{14}{\rm N}}$ binding energy of 6.27$\pm$0.27~MeV 
(E07 event IRRAWADDY, Fig.~\ref{fig:14N}). It would require an abnormally 
strong repulsive $\Xi NN$ three-body force to accommodate both~\cite{FG25}. 
A likely resolution of such incompatibility is that the 
$^{14}$N $\Xi^-_{1s}$-assigned capture events stand for $\Xi^0_{1p}-{^{14}
{\rm C}}$ nuclear states~\cite{FG23} briefly discussed below. 

\begin{figure}[!ht]
\begin{center}
\includegraphics[scale=0.55]{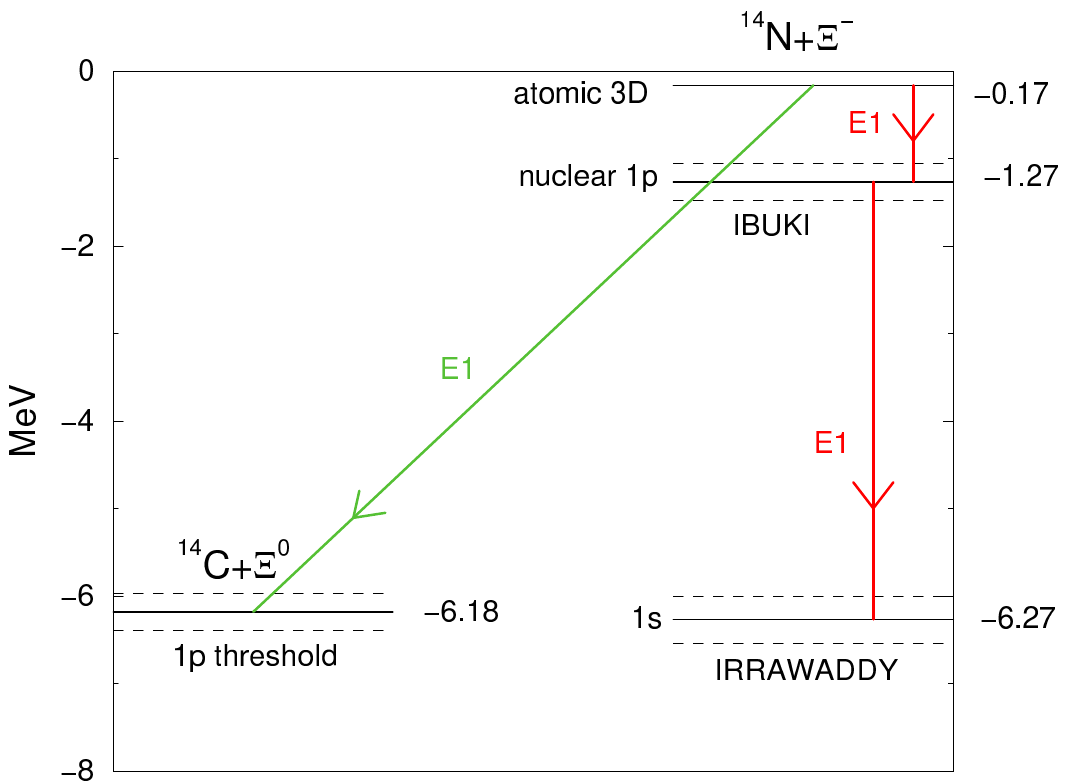} 
\caption{$\Xi^-_{1s}-{^{14}{\rm N}}$ IRRAWADDY reinterpreted as 
$\Xi^0_{1p}-{^{14}{\rm C}}$. Figure adapted from Ref.~\cite{FG23}.}
\label{fig:Xi0} 
\end{center} 
\end{figure}

$\Xi^-$ atomic cascade in light (C,N,O) emulsion nuclei normally terminates 
with atomic $3D\to 2P$ radiative $E1$ transitions~\cite{BFG99}. Indeed, 
$\Xi^-$ capture events identified during the last three decades at KEK and 
J-PARC provide evidence, as discussed in Ref.~\cite{FG21}, for $\Xi^-_{1p}$ 
nuclear states bound by $\approx$1~MeV in $^{12}$C and $^{14}$N. The formed 
$\Xi^-_{1p}$ Coulomb-assisted nuclear states undergo $\Xi^-p\to\Lambda\Lambda$ 
strong-interaction conversion. Forming radiatively $\Xi^-_{1s}$ nuclear bound 
states is suppressed by almost two orders of magnitude~\cite{Zhu,Koike}. 
Hence, recent $\Xi^-_{1s}$ assignments of capture events in $^{14}$N by J-PARC 
E07~\cite{E07} are questionable. Ref.~\cite{FG23} demonstrates how radiative 
$E1$ de-excitation of the atomic $\Xi^-_{3D}-{^{14}{\rm N}}$ state populates 
comparably both $\Xi^-_{1p}-{^{14}{\rm N}}$ and $\Xi^0_{1p}-{^{14}{\rm C}}$ 
states that are separated by about 5 MeV, see Fig.~\ref{fig:Xi0}, in spite of 
their rather small mixing by a residual two-body $\Xi N$ interaction of order 
0.5 MeV. This small mixing gets compensated by a roughly six times larger 
$E1$ transition energy from $\Xi^-_{3D}-{^{14}{\rm N}}$ to $\Xi^0_{1p}-{^{14}
{\rm C}}$ than to $\Xi^-_{1p}-{^{14}{\rm N}}$, resulting in two orders of 
magnitude enhancement of the $E1$ transition rate to $\Xi^0_{1p}-{^{14}
{\rm C}}$ which compensates for the roughly 1\% mixing strength. 
Such mechanism does not work for the carbon or oxygen nuclear constituents of 
the (C,N,O) emulsion. 

In conclusion, the only remaining $\Xi^-_{1s}$ nuclear candidate is the 
J-PARC E05 $\Xi^-_{1s}-{^{11}{\rm B}}$ state bound by $B_{\Xi^-}=8.9\pm 
1.4{^{+3.8}_{-3.1}}$~MeV~\cite{E05}. It is compatible with $\Xi^-_{1p}$ 
Coulomb-assisted states bound by $\approx$1~MeV in $^{12}$C and $^{14}$N upon 
using a Pauli-corrected density-dependent $\Xi$-nuclear attractive potential 
of $\approx$21~MeV depth at nuclear-matter density $\rho_0=0.17$~fm$^{-3}
$~\cite{FG25}. Such a potential depth is within the range of values 
17$\pm$6~MeV derived in Ref.~\cite{HH21} from a $K^+$ quasi-free spectrum 
measured by BNL-AGS E906 in ($K^-,K^+$) on $^9$Be~\cite{E906} covering 
also lower values than ours deduced from KEK-E224~\cite{E224} and 
BNL-AGS E885~\cite{E885} ($K^-,K^+$) experiments on $^{12}$C targets. 
However, a depth value of 21~MeV is considerably larger than reported 
in recent theoretical evaluations, e.g., HAL-QCD~\cite{Inoue19}: 4~MeV; 
$\chi$EFT$@$NLO~\cite{Kohno19}: 9~MeV; SU(6)~\cite{GC21}: 6~MeV. 
A notable exception is provided by versions ESC16*(A,B) of the latest 
Nijmegen extended-soft-core $\Xi N$ interaction model~\cite{Nij20} where 
$\Xi$-nuclear potential depths larger than 20~MeV are derived, but then 
reduced substantially by $\Xi NN$ three-body contributions within the same 
ESC16* model.

\section{Implications of a recent J-PARC E07 ${_{\Lambda\Lambda}^{~13}}$B 
event} 

\begin{figure}[!h] 
\begin{center} 
\includegraphics[scale=2.0]{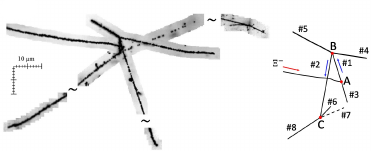} 
\caption{Photo of the new E07 event~\cite{He25} along with a schematic 
drawing of the ${_{\Lambda\Lambda}^{~13}}$B production vertex A and 
weak-decay vertices B and C; see text.} 
\label{fig:E07} 
\end{center} 
\end{figure}

J-PARC E07 published recently a new $\Xi^-$ capture emulsion event identified 
as production and decay of ${_{\Lambda\Lambda}^{~13}}$B \cite{He25} specified 
by the following three vertices (see Fig.~\ref{fig:E07}): 
\bigskip 
\begin{itemize}
\item
A:~~$\Xi^- + {^{14}{\rm N}}\to {_{\Lambda\Lambda}^{~13}{\rm B}}(\#1) + 
p(\#3) + n$
\item
B:~~$_{\Lambda\Lambda}^{~13}{\rm B}\to {_{\Lambda}^{5}{\rm He}}(\#2)+ t(\#4) 
+ {^3{\rm He}}(\#5) + 2n$
\item
C:~~$_{\Lambda}^{5}{\rm He}\to {^4{\rm He}}(\#6) + \pi^-(\#7) + p(\#8)$ 
\end{itemize}

We note that momentum and energy balance in both ${_{\Lambda\Lambda}^{~13}}$B 
production vertex A and weak-decay vertex B require unobserved neutrons. 
Vertex A yields $B_{\Lambda\Lambda}({_{\Lambda\Lambda}^{~13}}{\rm B})=25.57 
\pm 1.18\pm 0.07$~MeV, assuming here too that $\Xi^-$ capture occurs from a 
$^{14}$N $3D$ atomic state. 
Using $B_{\Lambda}({_{~\Lambda}^{12}}{\rm B}_{\rm g.s.})$=11.37~MeV, 
the $\Lambda\Lambda$ interaction strength in ${_{\Lambda\Lambda}^{~13}}$B 
assumes the value 
\begin{equation} 
\Delta B_{\Lambda\Lambda}=B_{\Lambda\Lambda}({_{\Lambda\Lambda}^{~13}}{\rm B})
-2B_{\Lambda}({_{~\Lambda}^{12}}{\rm B})=2.83\pm 1.18\pm 0.14~{\rm MeV}.   
\label{eq:DBLL13a}
\end{equation}  
Introducing the $(2J+1)$ avearged $B_{\Lambda}$ for the $(1^-,2^-)$ g.s. 
doublet levels instead of that for just the $1^-$ g.s. has a small effect of 
changing 2.83 MeV to 2.97 MeV, which is disregarded here. The value of 
$\Delta B_{\Lambda\Lambda}$ in Eq.~(\ref{eq:DBLL13a}) is considerably larger, 
by 2$\sigma$, than the value 0.67$\pm$0.17~MeV derived from the KEK-PS E373 
${_{\Lambda\Lambda}^{~~6}}{\rm He}$ NAGARA event~\cite{NAGARA2001,E373}. 

\bigskip 

J-PARC's E07 value of $B_{\Lambda\Lambda}({_{\Lambda\Lambda}^{~13}}{\rm B})$ 
disagrees with a value deduced by KEK-E176~\cite{E176} from a $\Xi^-$ capture 
event marked \# 15-03-37 in terms of a possible ${_{\Lambda\Lambda}^{~13}}$B 
production as follows: 
\begin{itemize} 
\item 
A:~~$\Xi^- + {^{14}{\rm N}}\to {_{\Lambda\Lambda}^{~13}}{\rm B} + p + n$ 
\item 
B:~~${_{\Lambda\Lambda}^{~13}{\rm B}}\to {_{~\Lambda}^{13}}{\rm C}^{\ast} + 
{\pi}^-$ 
\item 
C:~~${_{~\Lambda}^{13}}{\rm C}\to {^4{\rm He}} + {^4{\rm He}} + {^3{\rm He}} 
+ 2n$ 
\end{itemize} 
Here ${_{\Lambda}^{13}}{\rm C}^{\ast}$ stands for ${_{~\Lambda}^{13}}$C first 
excited doublet levels $J^{\pi}=(3/2^+,5/2^+)$ at $E_x\approx 4.9$~MeV. Vertex 
B then leads to $B_{\Lambda\Lambda}({_{\Lambda\Lambda}^{~13}}{\rm B})=23.6\pm 
0.8$~MeV for $B_\Lambda({_{~\Lambda}^{13}}{\rm C})= 12.0$~MeV~\cite{GHM16}, 
so that 
\begin{equation} 
\Delta B_{\Lambda\Lambda}=B_{\Lambda\Lambda}({_{\Lambda\Lambda}^{~13}}{\rm B})
-2B_{\Lambda}({_{~\Lambda}^{12}}{\rm B})=0.9\pm 0.8~{\rm MeV}, 
\label{eq:DBLL13b}
\end{equation}
consistently with the NAGARA event value. Vertex A confirms this result, 
but within a considerably larger uncertainty. 

Regarding the NAGARA event~\cite{NAGARA2001,E373} we note that 
(i) none of its vertices A and B 
\begin{itemize} 
\item 
A:~~$\Xi^- + {^{12}{\rm C}}\to {_{\Lambda\Lambda}^{~~6}{\rm He}} + {^3{\rm H}} 
+ {^4{\rm He}}$ 
\item 
B:~~$_{\Lambda\Lambda}^{~~6}{\rm He}\to {_{\Lambda}^{5}{\rm He}} + p + \pi^-$ 
\item
C:~~$_{\Lambda}^{5}{\rm He}\to p + d + 2n$
\end{itemize} 
requires unobserved neutrons; and (ii) neither ${_{\Lambda\Lambda}^{~~6}}$He 
nor ${_{\Lambda}^{5}}$He have particle-stable excited states that could modify 
$\Delta B_{\Lambda\Lambda}=0.67\pm0.17$~MeV. This makes NAGARA a {\it unique} 
event. 

A second-best case is ${_{\Lambda\Lambda}^{~10}}$Be with two observed events 
involving excited states. One event involves a $\Xi^-$-capture formation of 
${_{\Lambda\Lambda}^{~10}}{{\rm Be}^{\ast}_{2^+}}$(3 MeV)~\cite{DemYan2001}, 
whereas in the other event ${_{\Lambda\Lambda}^{~10}}{{\rm Be}_{\rm g.s.}}$ 
is formed, decaying weakly to ${_{\Lambda}^{9}}{{\rm Be}^{\ast}}$(3 MeV) 
by emitting a $\pi^-$ meson~\cite{Danysz63,Dalitz89}. No unobserved neutrons 
need to be invoked. Both derived values of $\Delta B_{\Lambda\Lambda}$ 
agree within experimental uncertainties, and also with the NAGARA value. 
Any different choice of excited-state pairs would lead to a substantial 
disagreement between the two values of $\Delta B_{\Lambda\Lambda}$ that follow 
from such a choice. No other $\Lambda\Lambda$ hypernuclear candidate beyond 
${_{\Lambda\Lambda}^{~~6}}$He offers this kind of unique determination of 
$\Delta B_{\Lambda\Lambda}$. This is also evident in both cluster-model 
calculations~\cite{Hiyama02,Hiyama10} and shell-model calculations 
\cite{GalMil11,GalMil12}.


\section*{Acknowledgments} 

Special thanks are due to Eliahu Friedman with whom I coauthored most of the 
works reviewed here in the first section. I would like also to thank the 
organizers of HYP2025 for their remarkable hospitality during the conference.

\end{document}